\documentclass[aps,prl,twocolumn,groupedaddress,showkeys]{revtex4}
\usepackage{epsfig}

\begin{document}


\title{Calculation of free energy landscapes: A Histogram Reweighted Metadynamics approach}

\author{Jens Smiatek}
\thanks{}
\email{jens.smiatek@uni-muenster.de}
\author{Andreas Heuer}
\thanks{}
\email{andheuer@uni-muenster.de}
\affiliation{Institut f{\"u}r Physikalische Chemie, Westf{\"a}lische Wilhelms-Universit{\"a}t M{\"u}nster, D-48149 M{\"u}nster, Germany}

\begin{abstract}
We present an efficient method for the calculation of free energy landscapes. Our approach involves a history dependent bias potential which is evaluated on a grid.
The corresponding free energy landscape is constructed via a histogram reweighting procedure a posteriori.
Due to the presence of the bias potential, it can be also used to accelerate rare events.
In addition, the calculated free energy landscape is not restricted to the actual choice of collective variables and can in principle be extended to auxiliary variables of interest
 without further numerical effort. The applicability is shown for several examples. 
We present numerical results for the alanine dipeptide and the Met-Enkephalin in explicit solution to illustrate our approach. Furthermore we derive an empirical formula that allows the prediction of the 
computational cost for the ordinary metadynamics variant in comparison to our approach which is validated by a dimensionless representation. 
\end{abstract}

\date{\today}

\maketitle

\section{Introduction}
\label{sec:introduction}
The characteristics and the behaviour of physical systems can be understood and predicted by the investigation of the underlying free energy landscape.
The knowledge of this often complex surface allows one to determine
reaction pathways for chemical reactions as well as stable configurations for proteins \cite{Dobson98,Onuchic97,Wales06,Wales06b,Dill08,Wolynes95} and glass-forming
systems \cite{Heuer08,Wales98,Walesbook}.
A successful tool for these studies are computer simulations which help to explore the characteristics of the system in detail.
Naturally, during the last years a lot of effort has been spent to develop novel efficient and time-saving methods.\\
Specifically in protein sciences, the free energy approach is promising for the investigation of unknown folding mechanisms.
To achieve a proper description of the dynamical behaviour, the trajectories may be projected onto a set of well chosen collective variables \cite{Dobson98,Onuchic97} which allow to define an effective energy 
landscape.
These variables can be interpreted as effective reaction coordinates spanning the phase space respectively the free energy space of the system.
Well known collective variables are the number of native contacts, dihedral angles
and the essential eigenvectors or principal components of the protein \cite{Dobson98,Onuchic97}. Thus a set of well suited reaction coordinates offers the opportunity to
describe the folding or unfolding of a
protein adequately without too much loss of information.\\  
However, it has to be mentioned that in some cases suitable collective variables are hard to determine. The investigation of the underlying dynamics of the system can then be performed 
by computational methods like discrete path sampling \cite{Walesbook,Wales2002} and transition path sampling \cite{Dellago98,Dellago02} to overcome this problem. These techniques represent computational algorithms
which are not dependent on the usage of collective variables.\\
Experiments and computer simulations have shown that naturally occuring proteins often have free energy landscapes with global funneling properties \cite{Dobson98,Onuchic97,Wales06,Dill08,Walesbook,Wolynes95}.
Nevertheless local minima can occur which represent trapped conformations of the protein \cite{Dobson98,Wolynes95}.
The transitions between these minima which are called rare events occur on timescales which are typically not accessible in computer simulations.
Over the last years several numerical methods have been developed to accelarate rare events and to compute the underlying free energy landscapes. In the following we
will mention the most common ones which have been well established in the scientific community. \\
Often used methods are thermodynamic integration \cite{Kollman93,Carter89,Sprik89} and free energy perturbation \cite{Bash87}. These techniques are well suited to calculate static properties like hydration
energies but normally fail in the calculation of free energy differences between specific folding mechanisms. To overcome this situation and to gain insight into
dynamical quantities, more sophisticated
ideas have to be adopted.\\
The umbrella sampling method \cite{Patey75} allows to determine free energy differences via the corresponding probability distributions of the whole accessible phase space.
An effective additional biasing potential drives
the system to more unlikely regions such that the whole phase space is visited. This idea has also been used
in adaptive force biased calculations \cite{Darve01,Darve08}, steered as well as adiabatic Molecular Dynamics simulations \cite{Gulligsrud99, Rosso02} and multicanonical
sampling approaches \cite{Berg91, Berg92, Mitsutake01}. Another promising method for the investigation of folding properties is the replica-exchange algorithm \cite{Earl05}.\\
Specifically in the last years several techniques have been proposed which employ an adaptively varying bias potential as an estimate for the free energy.
Examples for these methods are the Metadynamics algorithm in its several variants \cite{Laio02,Laio08} and the related
local elevation method respectively the conformational flooding approach \cite{vanGunsteren94,Grub95}.\\
The main idea of the Metadynamics method relies on a successive flattening of the free energy landscape by an additional potential energy in form of small gaussian hills
which are positioned at relaxed times at the present location of the
system. Then the free energy can be estimated as a negative mirror of the
potential energy which is exact in the limit of long simulation times \cite{Laio08}.
It has been shown that this method is in principle valid for all collective
variables and offers a broad range of applicability \cite{Laio08}. Even the
calculation of microscopic averages within the metadynamics scheme has been
successfully examined \cite{Tiana08}.
Nevertheless the drastic error dependence
on the parameters of the algorithm \cite{Laio05,Bussi06} in the original metadynamics scheme has to be mentioned. 
Keeping the errors
within tolerable ranges results in a drastic increase of simulation time.
To avoid local bumps in the landscape and to keep the errors low,
extensions of the existing Metadynamics method
\cite{Barducci08,Bonomi09,Barducci10,Micheletti04} 
have been proposed which offer a more reliable derivation.\\
In this paper we present a Histogram reweighted Metadynamics technique which offers the opportunity to overcome some drawbacks of the original method and its variants.
Our method is grid-based which means that the energy landscape spanned by the collective variables is divided into several regions.
By adding a short range cut-off biasing potential which is only evaluated at
the grid points of a predefined grid, the corresponding simulation time can be massively decreased.
Finally, the corresponding biased probability distributions are reweighted by histogram techniques \cite{Ferrenberg88,Kumar92,Roux95,Dill05} to compute the free energy landscape
a posteriori. It has been noticed that these techniques offer a low error dependent description \cite{Roux95,Dill05}.\\
Additionally a projection scheme is introduced which allows the investigation of further
collective variables if the probability distributions of the studied variables are well overlapping and if the projected coordinates are well-suited and fastly varying.
Although it is clearly impossible to reconstruct the whole highly dimensional phase space of the protein, our projection scheme allows to investigate additional low dimensional reaction coordinates like dihedral angles 
or spatial distributions in good agreement to directly derived energy landscapes.
As a consequence of this scheme, no further simulation time is required as the analysis can be performed a posteriori and the landscape is not restricted to the actual chosen set of coordinates.\\
This becomes useful by regarding the fact
that difficulties for predefined collective variables may arise, if they are
not the true reaction coordinates of the system \cite{Karplus04, Laio08} or give unsuitable descriptions of the corresponding free energy landscapes.
This is in particular a problem if the underlying folding or unfolding mechanisms are too complex for being projected onto
a low dimensional subspace which may lead to wrong conclusions \cite{Laio08} or if the phase space cannot be sampled efficiently. 
In these cases, discrete path sampling \cite{Walesbook,Wales2002}, transition path sampling \cite{Dellago98,Dellago02} and additional methods \cite{Karplus04,Muff2008,Noe2008,Strodel2008}
allow to investigate the hidden complexity of the free energy landscape without the usage of collective variables as alternative approaches.\\ 
By comparison to other approaches, it has to be noticed that in recent publications \cite{Bussi06,Ensing04,Babin06} the usage of umbrella sampled biased
probability distributions respectively using histogram reweighting procedures \cite{Marinelli09} 
for the correction of free energy landscapes has also been claimed. 
Even the evaluation of the underlying potential on a grid point has been
recently proposed for metadynamics \cite{Babin06} as well as for a closely related adaptive bias molecular dynamics scheme \cite{Babin08}. Both ingredients have also been applied
in a grid-based adaptive umbrella sampling scheme \cite{Bartels97} over a
decade ago and the above mentioned techniques are implemented in common software packages \cite{Plumed09,Plumed,NAMD10,NAMD,Amber11,Woods2005}.\\ 
A main reason for the introduction of grid evaluations is a massive decrease in computation time which scales by a constant factor. 
This fact is also used in
\cite{Babin06} where smoothly truncated Gaussians are evaluated by a kd-tree. 
Further realizations of grid techniques incorporate adaptive grids where the potential energy is calculated by a polynomial extrapolation on the grid points \cite{Plumed,NAMD},
choosing the potential of a close grid point as the biasing potential on the particle \cite{Plumed} or applying cut-off radii for the gaussians \cite{Plumed}.
It was also shown that the above mentioned adaptive biased grid approach in its computational cost scales linearly with simulation time in contrast to 
ordinary metadynamics \cite{Babin08}. Here cubic B-splines are used for the evaluation of the biasing potential on the corresponding grid points.\\
Our approach has the advantage of a well-defined simple potential all over the energy surface. This allows to change the resolution of the grid on-the-fly even
after the simulations have finished to
resolve regions of specific interest in more detail. In addition, it is easy
to implement and scales with a constant number $\mathcal{O}(2 d)$ of calculations per
timestep, where
$d$ is the number of collective variables, respectively the dimension of the
grid.\\ 
As a further remark, our potential energy landscape is purely used as a biasing potential to accelerate
rare events. Hence, the fine resolution of the free energy landscape is
achieved by histogram techniques. This allows us to use a very rough potential energy surface created in a short simulation time. 
Additionally the bias force is always well-defined such that discontinouities are avoided.\\ 
Therefore our method in its
interpretation is closely related to the Wang-Landau approach \cite{Wang01} and adaptive
umbrella biasing techniques \cite{Bartels97} with the effectiveness of \cite{Babin08}. 
This finally results in a simple and robust algorithm which is easy to
implement in common public software packages like GROMACS
\cite{Berendsen95,Hess08,Spoel05} or other programs 
and allows to tune the resolution of the landscape on demand easily after the
simulations have finished.\\ 
We further derived an empirical formula for the computational cost required for the ordinary metadynamics scheme in comparison to our approach. Thus we were able 
to show that the grid technique for reasons of computational efficiency is often more preferable especially for a large number of hills in comparison to the number of atoms in the system.\\ 
The paper is organized as follows. In the second section we introduce the method and the theoretical background. The third and the fourth section illustrate model
test cases and the numerical details of the following peptide simulations.
The fifth section presents the results for the alanine
dipeptide and the Met-Enkephalin. In the sixth section the values for the computational cost required for both methods are investigated and an empirical formula is derived.
We conclude with a brief summary in the last section.
\section{Histogram Reweighted Metadynamics: Theoretical background}
\label{sec:theory}
The system we consider is described by a set of coordinates $x$ evolving under the action of dynamics following the trajectory $x(t)$ and described by a canonical equilibrium distribution at temperature $T$. The set of
coordinates $x$ may include atomic positions or angles as well as any other auxiliary collective variable representing the characteristics of the system.\\
If the system shows metastability, some regions separated by large energy barriers cannot be explored by the evolution of the trajectories in a reasonable simulation time.
Guided by the conventional metadynamics approach an additional potential has to be added at specific constant times $t_1, t_2,...t_{_{N}}$ on the
trajectory $x(t)$ to overcome the barriers and to accelerate the rare events. It has to be ensured that the protein relaxes between these times such that
the system diffuses in the next local minimum.
For the force evaluation in the ordinary Metadynamics method, the sum over all previously added gaussian functions has to be performed \cite{Laio02} which results in a strong increase
of computation time.
To avoid this increase,
we use an approach where the additional potential energy is evaluated 
on the grid points of a predefined grid closely related to \cite{Bartels97,Babin08,Babin06}, spanning the whole range of the accessible phase space in the set of collective variables $x$.\\
The grid points $x_{_{G,i}}$ are separated by the grid constant $\sigma_{_G}$,
which is the distance between two neighboring grid points in one dimension.
The system now evolves in time moving over the energy landscape covered by the grid. At the times $t=t_1,t_2,...t_{_{N}}$ the following potential energy
\begin{equation}
\label{eq:spread}
\Delta V_{T}(x_{_{G,i}},t)=\mathcal{G}(x-x_{_{G,i}})\;\omega
\end{equation}
with
\begin{equation}
\label{eq:weight}
\mathcal{G}(x-x_{_{G,i}})=e^{-\frac{(x(t)-x_{_{G,i}})^2}{2 x_{_c}^2}}\left(1-\left(\frac{|x(t)-x_{_{G,i}}|}{x_{_{c}}}\right)\right)
\end{equation}
defined by
\begin{eqnarray}
\begin{array}{cc}
\mathcal{G}(x-x_{_{G,i}})
\neq 0 & : |x(t)-x_{_{G,i}}| < x_c \\
\mathcal{G}(x-x_{_{G,i}})=0 & : |x(t)-x_{_{G,i}}|\geq x_c
\end{array}
\end{eqnarray}
is evaluated on all grid points $N_{_{GK}}$ whose distance from $x(t)$ is closer than the cut-off radius $x_c$.
The magnitude of the potential energy is given by
the height $\omega$ like in the conventional metadynamics scheme. As it has been
discussed in \cite{Laio08} the values for $\omega$ have to be low.\\  
The potential energy at the grid points then evolves in simulation time with
\begin{equation}
\label{eq:hist}
V_{T}(x_{_{G,i}},t_{_{N}})=V_{T}(x_{_{G,i}},t_{_{N-1}})+\Delta V_{T}(x_{_{G,i}},t_{_{N}})
\end{equation}
emerging rapidly if the specific neighborhood of the grid point is often visited by the trajectory, {\em e.g.} in free energy minima.
Then we need to update $2d$ values of grid points for each calculation, where $d$ is the number
of collective variables or in other words the dimension of the grid.\\
The biasing potential exerted on the system at times $t\geq t_{_{N}}$ yields
\begin{equation}
\label{eq:bias}
V_B(x,t) = \sum_i^{N_{GK}}\mathcal{G}(x-x_{_{G,i}})\;V_T(x_{_{G,i}},t_{_{N}})
\end{equation}
where the summation is over the number of grid points $N_{GK}$ within $x_c$,
with the actual value of the potential $V_{T}(x_{_{G,i}}, t_{_{N}})$ of each grid point at $x_{_{G,i}}$ as defined in Eqn.~(\ref{eq:hist}) and
the weighting factor $\mathcal{G}(x-x_{_{G,i}})$ of Eqn.~(\ref{eq:weight}).
The resulting local minimum between two grid points can usually be neglected if the distance between neighboring grid points $\sigma_G$
is small
compared to the typical expected diameter of a free energy minimum, respectively a high resolution of the grid.
Repeating the whole scheme allows to fill the minima efficiently until the potential energy is reached to overcome the energetic barriers. The energy
landscape with constant values $V_T(x_{_{G,i}})$ and the corresponding continuous function $V_B(x)$ are finally used in the biased simulation runs to produce flat
probability distributions.
\begin{figure}
\includegraphics[width=0.5\textwidth]{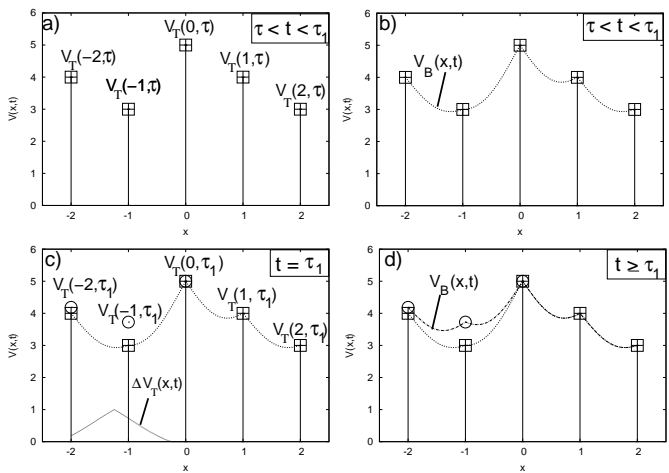}
\caption{
    \label{fig:one} Illustration of the Histogram Reweighted Metadynamics scheme with $x_c=\sigma_{_G}=1\sigma$. 
Panel A shows the constant potential energy $V_T(x_{_{G,i}},\tau)$ of the grid points at times $\tau<t<\tau_1$. The biasing potential energy $V_B(x,t)$ of 
Eqn.~(\ref{eq:bias}) for times $\tau<t<\tau_1$ is shown in Panel B. At times $t=\tau_1$ (Panel C) a new gaussian function $\Delta V_T(x,t)$ is placed at the actual 
position of 
the system at $x=-1.25\sigma$ which is only evaluated at the grid points such that $\Delta V_T(x_{_{G,i}})$ is summed to 
the old potential $V_T(x_{_{G,i}},\tau)$ via Eqn.~(\ref{eq:hist}) yielding the new values $V_T(x_{_{G,i}},\tau_1)$(shown as circles).
The resulting new biasing potential function $V_B(x,t)$ is shown in Panel D. 
}
\end{figure}
An illustration of the scheme is presented in Fig.~\ref{fig:one}.\\
It has to be noticed that the specific choice of the biasing potential is
arbitrary \cite{Laio08,Babin06} although in combination with the grid potential evaluation it
can be pointed out that the function proposed in Eqn.~(\ref{eq:weight}) offers the advantage of a fast computation. Due to the well-defined
potential at every position, it can be even used to achieve a finer resolution of the
grid in combination with histogram
reweighting techniques.
As a further remark, it has to be mentioned
that the final biased simulations as well as the previous simulation runs for the construction of the
biasing potential can be trivially parallelized in the spirit of the multiple
walker metadynamics technique \cite{Raiteri06} to save computation time.\\ 
In the following we give a short description of the histogram reweighting technique.
The free energy landscape is finally calculated 
via the WHAM equations \cite{Ferrenberg88, Kumar92, Roux95, Dill05} 
by a reweighting of the biased probability distributions. 
This procedure allows
the calculation of the free energy within a finite tolerance value \cite{Dill05} and its combination to the biasing potential for 
refinement and construction has been shown to be advantageous \cite{Laio08,Laio05,Babin06,Babin2010}.
The scheme is given by two equations
\begin{equation}
\label{eq:pfree}
P(x) = \frac{\sum_{i=1}^{N_{sim}}n_i(x)}{\sum_{i=1}^{N_{sim}}N_i e^{-(F_i-V_B(x))/k_BT}}
\end{equation}
and
\begin{equation}
\label{eq:ffree}
   F_i =-k_BT \log\left(\sum_{x_{bins}}P(x) e^{-V_B(x)/k_BT}\right)
\end{equation}
which allow to compute the best estimate of the unbiased probability $P(x)$ where $N_{sim}$ denotes the number of independent simulations, $n_i(x)$ the number of counts
in a histogram bin associated with $x$,
$V_B(x)$ is the final constant biasing potential of Eqn.~(\ref{eq:bias}) at position $x$ and the free energy shift is $F_i$ for each simulation with thermal energy $k_BT$.
Due to the fact that $F_i$ and $P(x)$ are unknowns, Eqn.~(\ref{eq:pfree}) and Eqn.~(\ref{eq:ffree}) have to
be solved by iteration to self-consistency within a predefined tolerance value. The values of $P(x)$ can then be used to calculate the resulting free energy of the bin via
\begin{equation}
\label{eq:freenrg}
F(x)= -RT\log\left(\frac{P(x)}{P(0)}\right)
\end{equation}
where $P(0)$ is a reference value and $R$ denotes the molar gas constant \cite{Roux95}. It has to be pointed out, that the error in the energy is strongly dependent on the resolution of the histogram leading 
to more or less pronounced approximations. 
For a detailed description of the method we refer the reader to Refs. \cite{Kumar92,Roux95}.\\
As it was mentioned in the introduction, the main part of a successful calculation relies on the choice of appropriate collective variables on which the free energy surface is spanned.
Well-suited collective variables are dihedral angles as well as the essential eigenvectors of the system
\cite{Amadei93}. It has been recently demonstrated that the application of eigenvectors as collective variables results in adequate descriptions of the folding mechanisms
and of the
free energy landscape in ordinary Metadynamics computations \cite{Spiwok07,Spiwok08}.
Thus we follow this approach due to the assumption that nearly all relevant motion is captured in the
first eigenvectors of the system \cite{Amadei93}. This is remarkable by regarding
the fact that drastic difficulties appear if wrong reaction
coordinates are chosen which do not capture the main motion \cite{Laio08, Karplus04}. But nevertheless, it has again to be remarked that the usage of eigenvectors 
is not a guarantee for correctness of the free energy landscape due to hidden complexities in higher dimensional collective variables \cite{Laio08}.
In the following we give a brief description of the method.\\
The essential eigenvectors can be calculated by the superimposed coordinates $\vec{r}$ of $N$ atoms of the system which build the covariance matrix ${\bf C}$ via
\begin{equation}
{\bf C}=<(\vec{r}-<\vec{r}>)(\vec{r}-<\vec{r}>)^T>
\end{equation}
where $i,j = 1,2,\dots 3N$ and the time-averaged or reference value is denoted by the brackets $<\ldots>$. The diagonalization of ${\bf C}$ leads to
\begin{equation}
\bf{C}={\bf E}\Lambda{\bf E}^{-1}
\end{equation}
where ${\bf E}$ is a matrix of eigenvectors and ${\bf \Lambda}$ is a matrix of eigenvalues marking the positional fluctuations.
Sorting the eigenvalues in decreasing order allows to identify the largest positional fluctuations with all important structural transitions by the first eigenvectors which
form the essential dynamics of the system \cite{Amadei93}.
The projection at time $t$ on the $i-$th eigenvector is then defined by
\begin{equation}
p_i(t)=(\vec{r}(t)-<\vec{r}>)\cdot \vec{e}_i
\end{equation}
with the specific eigenvectors ranging from $i=1,2\dots 3N$.\\
If the motion of the system is not restricted to the essential subspace in the biased simulations such that even lower eigenvectors contribute,
the potential biasing energy of Eqn.~(\ref{eq:bias}) even activates the motion of the remaining subspace. Hence the concerted motion of all eigenvectors
is influenced by the biasing energy. Therefore a projection scheme allows to construct the free energy landscapes of additional collective variables like lower
eigenvectors or dihedral angles a posteriori.\\
The main idea is that the applied bias potential energy drives the system through structural transitions which could
be in principle also observed in additional collective variables.
The occurrence of the variables $x(t)$ have to be projected
onto the new collective variable $\xi(t)$ exposed at the same time.
Formally, this is given by
\begin{equation}
f\;:\; x(t)\rightarrow\xi(t),
\end{equation}
together with the corresponding biasing potential energy
\begin{equation}
f\;:\;V_B(x,t)\rightarrow V_B(\xi,t).
\end{equation}
After the final timestep $\tau$, the constant potential energy at the new grid points $\xi_{_{G,i}}$ can be evaluated
by using the maximum last values of $V_B(\xi,\tau)$. Choosing the maximum value for $V_B(\xi_{_{G,i}})$ within certain small regions $\delta$, where $\delta$ is the
half distance between two grid points gives
\begin{equation}
V_B(\xi_{_{G,i}})=\max(V_B(\xi_{_{G,i}}\pm\delta,\tau))
\label{eq:proj}
\end{equation}
for the construction of the potential energy landscape in the new set of chosen reaction coordinates. The values for $V_B(\xi_{_{G,i}})$ can now be inserted into
the Eqns.~(\ref{eq:pfree}) and (\ref{eq:ffree}) with the corresponding biased probabilities $P(\xi_{_{G,i}})$.
The final potential energy landscape in the new set of
collective variables of Eqn.~(\ref{eq:proj}) is evaluated by the existing data
and by explicit calculation 
of each $\xi(t)$.
Sampling the biased probabilities in the new set of variables finally allows
to determine the unbiased probabilities by using the histogram reweighting
procedure in the new set of collective variables at any resolution.\\
Again it has to be ensured that the motion is not constrained and that the potential energy even activates lower eigenvectors to establish a free system behaviour.
In principle all collective variables which are fastly varying and cover a small subspace 
can be seen as suitable reaction coordinates if a sufficient simulation time is given.
Additionally they also have to clearly distinguish between different states of conformations, can be well sampled meaning a good overlap in the distribution functions and covering of the phase space 
and show no hysteresis effects due to hidden complexities \cite{Laio08}.
Several publications discuss this problem and new techniques to overcome it in more detail \cite{Walesbook,Wales2002,Dellago98,Dellago02,Karplus04,Muff2008,Noe2008,Strodel2008}.
If these requirements are fulfilled, the projected energy landscape into the new collective variables can be seen as reliable. Nevertheless, it has to be noticed that the validity of a landscape is related
to the suitability of the chosen projected collective variables. This is also true for the original chosen coordinates.
\section{1- and 2-Dimensional Model potential}
\label{sec:testcase}
The first test case for our approach is a particle confined in a one-dimensional well defined potential
\begin{equation}
\label{eq:test}
V(x) = \epsilon\left[\left(\frac{x}{\sigma}\right)^4-10\left(\frac{x}{\sigma}\right)^2\right]
\end{equation}
where $\epsilon$ is the unit thermal energy $k_BT$ and the length unit is given by $\sigma$. We performed a Monte Carlo simulation with the Metropolis criterion \cite{Smit} 
consisting of $10^6$ timesteps where every $500-$th step a potential energy with height $\omega=0.25\epsilon$ has been set. The grid points were separated by
a grid constant $\sigma_{_G}=0.1\sigma$. The cutoff radius for the construction of potential function has been chosen to $x_c=0.2\sigma=2\sigma_{_G}$ and
$x_c=\sigma_{_{G}}$ for the
evaluation of the biasing potential.
Having constructed the biasing energy landscape, we performed four simulations with $10^6$ timesteps to derive the biased probability distribution functions which
have been reweighted
by the WHAM equations (Eqns.~(\ref{eq:pfree}) and (\ref{eq:ffree})).
\begin{figure}                                                                                                                                                               
\includegraphics[width=0.5\textwidth]{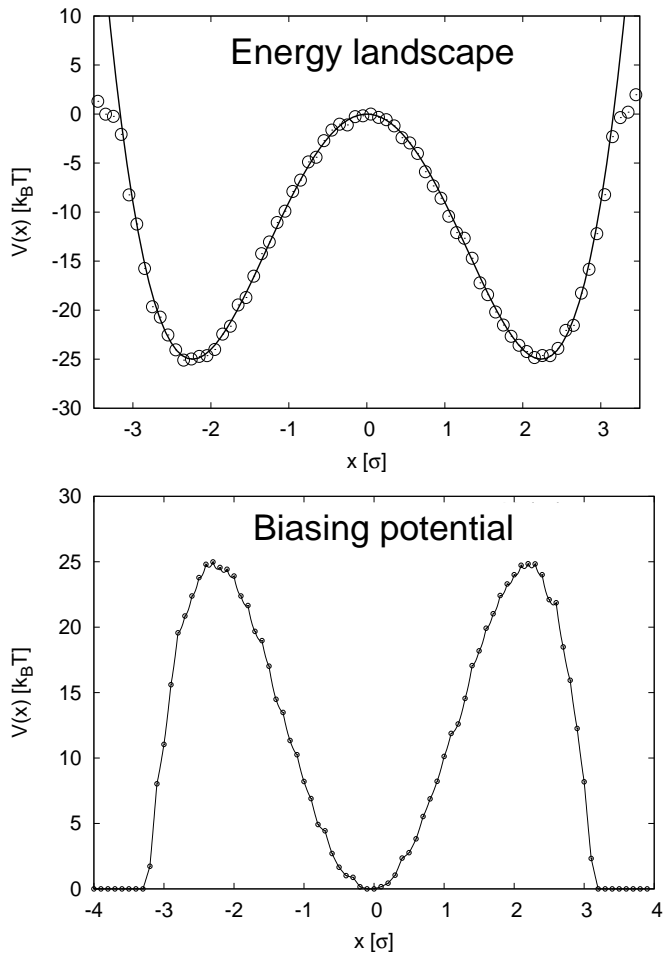}                                                                                                                 
\caption{\label{fig:two} {\bf Top:} Shifted energy landscape for the test potential of Eqn.~(\ref{eq:test}). 
The circles correspond to the values derived by the Histogram reweighted Metadynamincs approach whereas the black line corresponds to Eqn.~(\ref{eq:test}). 
The data points are in good agreement to Eqn.~(\ref{eq:test}) except for values $|x|>3.0\sigma$.          
This is due to the error of the statistical accuracy of the probability distribution which is caused by the rapid increase of the test potential.
{\bf Bottom:} Continous biasing potential $V_B(x)$ (solid line) which is used for the calculation of the biasing probabilities and the corresponding values of the potential energy $V_T(x_{_{G,i}})$ (points) 
evaluated at the grid points.}                                                                                                                                                                             
\end{figure}                                                        
Fig.~\ref{fig:two} presents the numerical results for the potential of Eqn.~(\ref{eq:test}). The numerical values are in good agreement to the theoretical results of the test potential except for
$|x|>3.0\sigma$. The reason for this misbehaviour can be related to the low
statistical accuracy of the probability distribution \cite{Parrinello05} which is caused by the rapid increase of the model potential at these points.
The constant potential $V_B(x)$ of Eqn.~(\ref{eq:bias}) as well as the corresponding discrete potential energy values of the grid points of Eqn.~(\ref{eq:hist}),
which have been used for the biased simulations are shown in the lower panel
of Fig.~\ref{fig:two}.
Nevertheless the results of this simple model potential
have shown that our method works and produces accurate results in one dimension.\\
Another 2-dimensional test potential is given by
\begin{equation}
V(x,y) = \epsilon\left[\left(\frac{x}{\sigma}\right)^4+\left(\frac{y}{\sigma}\right)^4-10\left(\left(\frac{x}{\sigma}\right)^2+\left(\frac{y}{\sigma}\right)^2\right)+50\right]
\label{eq:2d}
\end{equation}
which represents four energetic minima. We performed a couple of Langevin Dynamics simulations \cite{Schlick02}
obeying the Langevin equation which is given by
\begin{equation}
\vec{F}_i = -\zeta \vec{v}_i+\vec{\eta}_i
\end{equation}
where the force $\vec{F}_i$ on a particle depends on the friction coefficient $\zeta$, the velocity of the particle $\vec{v}_i$ and the stochastic force
$\vec{\eta}_i$. The stochastic force represents the thermal brownian motion of the particle with the following properties
\begin{equation}
<\vec{\eta}_i(t)> = 0
\end{equation}
and
\begin{equation}
<\eta_{i\alpha}(t)\eta_{j\beta}(t^{\prime})> = 2 \zeta k_BT \delta_{ij}\delta_{\alpha\beta} \delta(t-t^{\prime})
\end{equation}
which ensures the presence of the fluctuation dissipation relation. Thus the stochastic force is delta-correlated white noise which ensures a canonical ensemble
at thermal equilibrium. We chose $\zeta=1\sqrt{m\epsilon}/\sigma$ with the mass $m$, the temperature $T=1$ and ran a simulation of $10^7$ timesteps,
where every $1000$-th step a gaussian function has been set. We chose as values for the parameters
$\sigma_{_G}=0.25\sigma$, $x_c=0.5\sigma$, $\omega=0.1\epsilon$ and we used a timestep of $\delta t=0.01 \sigma/\sqrt{m\epsilon}$.
\begin{figure}
\includegraphics[width=0.5\textwidth]{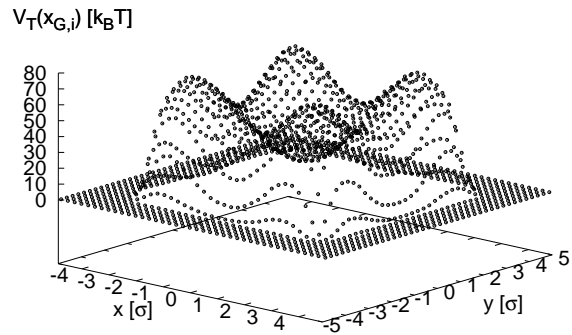}
\caption{\label{fig:gauss}
Values of $V_T(x_{G,i})$ which have been used in the biased simulations for the 2-dimensional test potential of Eqn.~(\ref{eq:2d}).
}
\end{figure}
The final values for $V_T(x_{G,i})$
of Eqn.~(\ref{eq:hist}) are shown in Fig.~\ref{fig:gauss}. These values have been used for the evaluation of the biasing potential of Eqn.~(\ref{eq:bias}).
\begin{figure}
\includegraphics[width=0.5\textwidth]{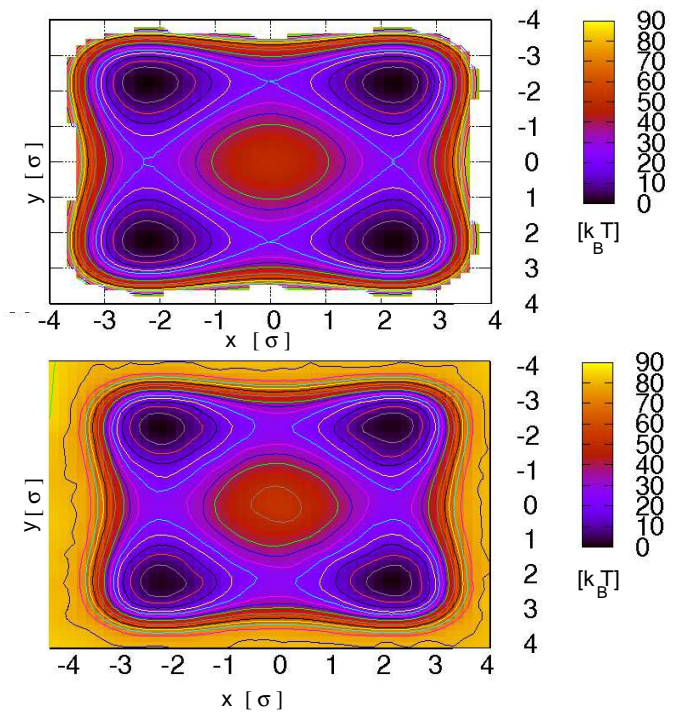}
\caption{\label{fig:nrg}
{\bf Top:} Energy profile of Eqn.~(\ref{eq:2d}). {\bf Bottom:} Resulting reweighted energy profile. The solid lines correspond to energy differences of $5 k_BT$.
}
\end{figure}
We performed
four simulation runs with $2.5\cdot 10^6$ timesteps where again every $1000$-th step the position of the particle has been stored for the sampling of the underlying free
energy landscape.
The resulting reweighted energy profile is presented in Fig.~\ref{fig:nrg}. The good agreement to the energy profile of Eqn.~(\ref{eq:2d}) is obvious. Larger deviations
only occur at positions $x,y > 3.5 \sigma$ above 60 $k_BT$ due to poor statistical accuracy. The standard
deviation to the correct profile is around 5\% per value.
\section{Numerical details}
All our Molecular Dynamics simulations have been performed by the software package GROMACS \cite{Berendsen95,Hess08,Spoel05} in which our in-house written code for the 
Histogram reweighted 
Metadynamics method has been implemented. The source code is available on request by contacting one of the authors.
\subsection{Alanine dipeptide}
We performed our Molecular Dynamics simulations by using the GROMACS ports of the AMBER94 force field \cite{Cornell95, Sorin05}.
The simulation box contains $313$ TIP3P water molecules \cite{Jorgensen83} within a dimension of $2.005\times 2.194\times 2.192$ nm. The time step was $1$ fs. The temperature
 $T=300$ K
was kept constant by a Nose-Hoover thermostat.
\begin{figure}
\includegraphics[width=0.35\textwidth]{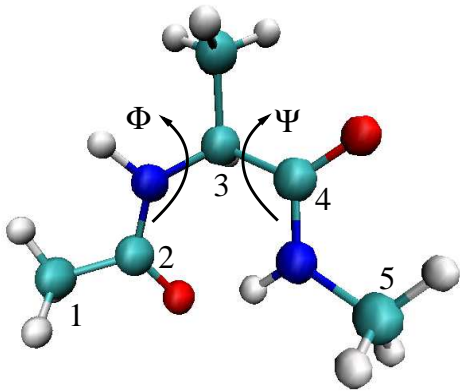}
\caption{
    \label{fig:three} Schematic representation of the alanine dipeptide with dihedral angles $\Phi$ and $\Psi$.
 The numbered C$_{\alpha}$ carbon atoms have been used for the eigenvector analysis.}
\end{figure}
Electrostatics have been calculated by the PME (Particle Mesh Ewald) method. All bonds have been constrained by the SHAKE algorithm \cite{SHAKE}.\\
After a short equilibration run of 350 ps we performed a 500 ps unbiased simulation to analyze the eigenvectors of the system constructed by the numbered C$_{\alpha}$ carbon
atoms of Fig.~\ref{fig:three}.
The corresponding potential energy landscape has been constructed in the phase space of the first two
eigenvectors within a 2 ns simulation run. The grid constant has been chosen to $\sigma_{_G}=0.01$ nm, the height of the potential function has been set to $\omega=0.2$
kJ/mol
and the relaxation time was $\tau=1$ ps.
The cut-off radius $x_c$ was $0.02$ nm for hill setting and $0.01$ nm for the evaluation of the biasing forces.
The final biased probability distributions have been derived by four several independent $2$ ns simulations with different temperatures $300,310,320$ and $330$ K whose
additional kinetic energy allows to explore the phase space more efficiently.
Nevertheless the WHAM equations
allow to combine the results for the different temperatures and give the results for a specific temperature if not too large deviations exist \cite{Roux95}.\\
For a comparison of the eigenvector free energy landscape we further applied a conventional metadynamics scheme as described as in \cite{Spiwok07,Spiwok08}. 
The simulation time was 2 ns where every picosecond a gaussian hill of width 0.01 nm 
with height 0.2 kJ/mol has been set. 
Additionally we used the software plug-in PLUMED \cite{Plumed09,Plumed} for a conventional metadynamics simulation to directly calculate the free energy landscape of the Ramachandran plot.
All parameters are identical to the histogram reweighted metadynamics scheme.
\subsection{Met-Enkephalin}
The molecule consists of five residues TYR-GLY-GLY-PHE-MET and its molecular structure has been taken from the PDB entry 1PLW \cite{Marcotte04}.
The force field was GROMOS96 \cite{Oostenbrink04}. The box size has been chosen to $3.214\times 3.214 \times 3.214$ nm with $1058$ SPC water molecules.
The time step was $2$ fs and the temperature
was kept constant by a Nose-Hoover thermostat. As mentioned above, all bonds have been constrained by the SHAKE algorithm \cite{SHAKE}.
Electrostatics have been again calculated by the PME (Particle Mesh Ewald) method.\\
After a warm up phase of $350$ ps we perfomed a $400$ ps simulation run for the analysis of the corresponding eigenvectors of all atoms at temperature $T=400$ K. We chose this high temperature to capture all
necessary transitions in the essential first eigenvectors.
The corresponding potential energy grid has been constructed in a 1 ns simulation run where the grid constant has been chosen to $\sigma_{_G}=0.01$ nm and the height of the potential function has been
set to $\omega=0.2$ kJ/mol with relaxation times of $\tau=2$ ps. The cut-off radius $x_c$ has been chosen to $0.2$ nm respectively $x_c=0.1$ nm as mentioned above.
The biased simulations consist of three independent runs of $T=300$ K and $T=305$ K with $4$ ns and $T=310$ K with $1$ ns simulation time. The corresponding timestep
was $\delta t = 2$ fs. All energy landscapes have been reweighted for a temperature $T=300$ K.
\section{Numerical results}
\label{sec:numerics}
\subsection{Alanine Dipeptide}
\begin{figure}
\includegraphics[width=0.5\textwidth]{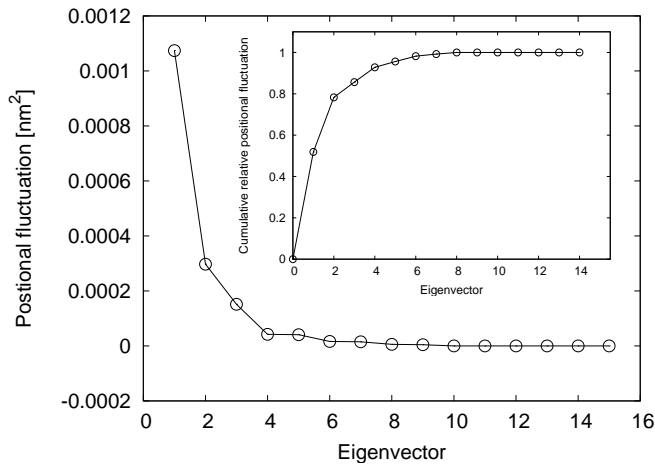}
\caption{
    \label{fig:four} Fluctuation and cumulative relative positional fluctuation (inset) of the eigenvectors. Nearly 80\% of the overall motion of the alanine dipeptide can be described
by the first two eigenvectors.}
\end{figure}

A well suited model system to test our approach is the alanine dipeptide (Fig.~\ref{fig:three}).
Alanine dipeptide is one of the most studied model systems over the last years \cite{Strodel2008,Smith99,Rao2007,Levy2004,Weinan2005,Dinner2005,Parrinello2007,Roux2008}.
The corresponding eigenvalues and the cumulative positional fluctuations are shown in Fig.~\ref{fig:four}. It is obvious that nearly 80\% of the overall positional fluctuation can be described
by the first two eigenvectors. Thus it can be assumed that all important structural transitions
should be captured by using these variables as reaction coordinates \cite{Amadei93}.\\
\begin{figure}
\includegraphics[width=0.5\textwidth]{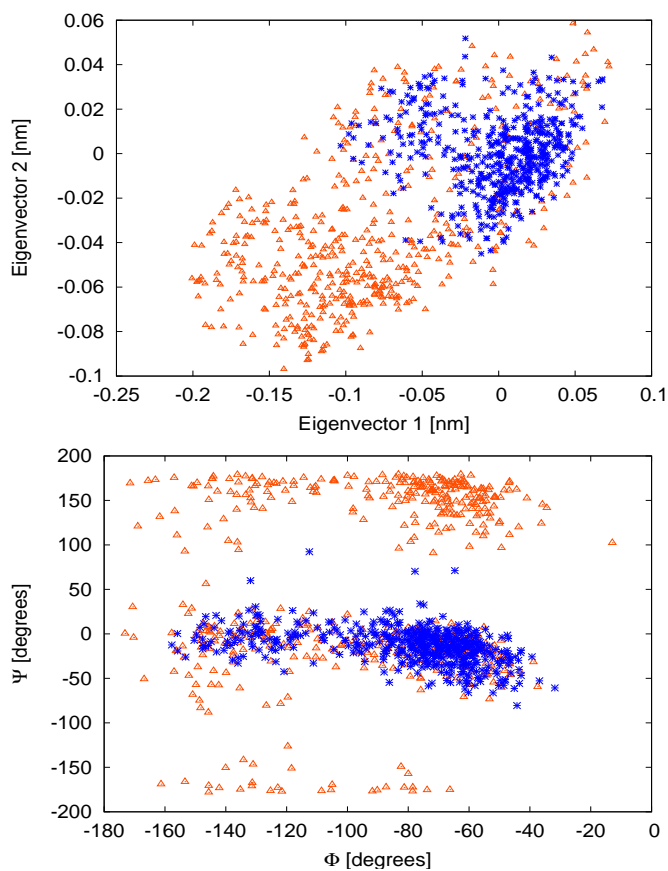}
\caption{
    \label{fig:five} {\bf Top:} Positions in the eigenvector space in the $300$ K biased simulation (orange triangles) and the unperturbed simulation results which are illustrated as blue stars. 
	{\bf Bottom:} Corresponding values for the dihedral angles $\Phi$ and $\Psi$ sampled from the same simulation runs.}
\end{figure}
The corresponding positions for the $300$ K biased simulations are presented in the top (orange triangles) of Fig.~\ref{fig:five} in contrast to an unperturbed simulation
run (blue stars) for the same parameter sets. The bottom figure illustrates the values for the dihedral angles $\Phi$ and $\Psi$ in the Ramachandran plot.
It is obvious that the biasing potential energy landscape drives the system into more unlikely minima to accelerate the rare events of these structural transitions.\\
\begin{figure}
\includegraphics[width=0.5\textwidth]{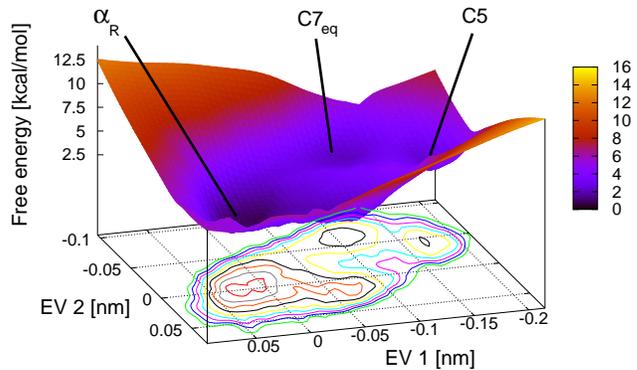}
\caption{
    \label{fig:six} Free energy landscape for the eigenvectors $1$ and $2$ of the alanine dipeptide. The lines correspond to energy differences of $0.5$ kcal/mol.}	
\end{figure}  
\begin{figure}
\includegraphics[width=0.5\textwidth]{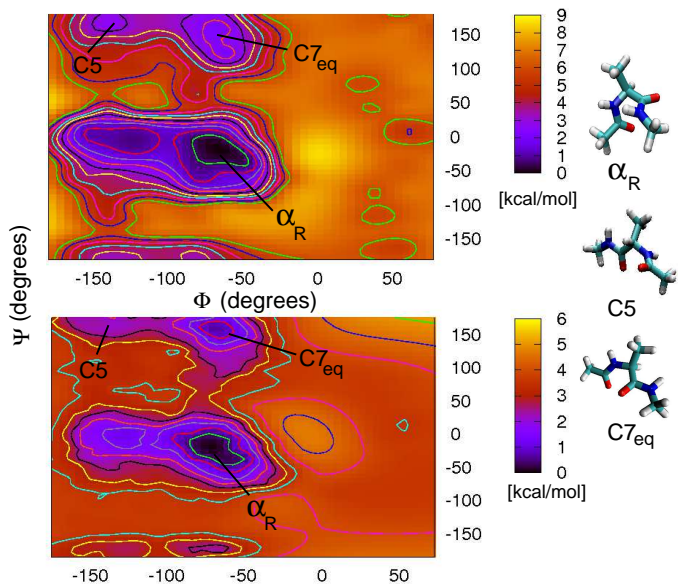}
\caption{
    \label{fig:seven} {\bf Top:} Free energy landscape derived by the projection of the eigenvectors 1 and 2 on the phase space of the dihedral angles $\Phi$ and $\Psi$. The marked regions denote stable configurations with the corresponding molecular representation.
        The lines correspond to energy differences of $0.5$ kcal/mol. {\bf Bottom:} Corresponding free energy landscape derived by a direct metadynamics simulation in the phase space of the 
corresponding dihedral angles $\Phi$ and $\Psi$.}
\end{figure}
The final free energy landscapes at $T=300$ K are shown in Figs.~\ref{fig:six} and \ref{fig:seven}.
Fig.~\ref{fig:six} presents the free energy landscape in the space of eigenvectors whereas Fig.~\ref{fig:seven} is the corresponding
Ramachandran plot in the space of the dihedral angles which has been derived by the proposed projection scheme and a direct conventional metadynamics simulation run in the phase space of the dihedral angles. 
Both figures illustrate the corresponding conformations of the alanine dipeptide and their relative
energy difference from the most stable configuration $\alpha_R$ to two further minima $C7_{eq}$ and $C5$.
The relative energy difference betweeen the $\alpha_R$ and the $C7_{eq}$ and $C5$ minima is given by $\Delta F \approx 1.9$ kcal/mol, respectively $2.2$ kcal/mol
which is in
good agreement to the results reported in Ref.~\cite{Spiwok07,Spiwok08} with $\approx$ 1.74 kcal/mol, respectively $\approx$ 2.10 kcal/mol. 
The good agreement in the location and the acceptable values of the free energy minima in comparison to a direct biasing between the plots shown in Fig.~\ref{fig:seven} are remarkable which validates
the proposed projection scheme. 
\begin{figure}
\includegraphics[width=0.5\textwidth]{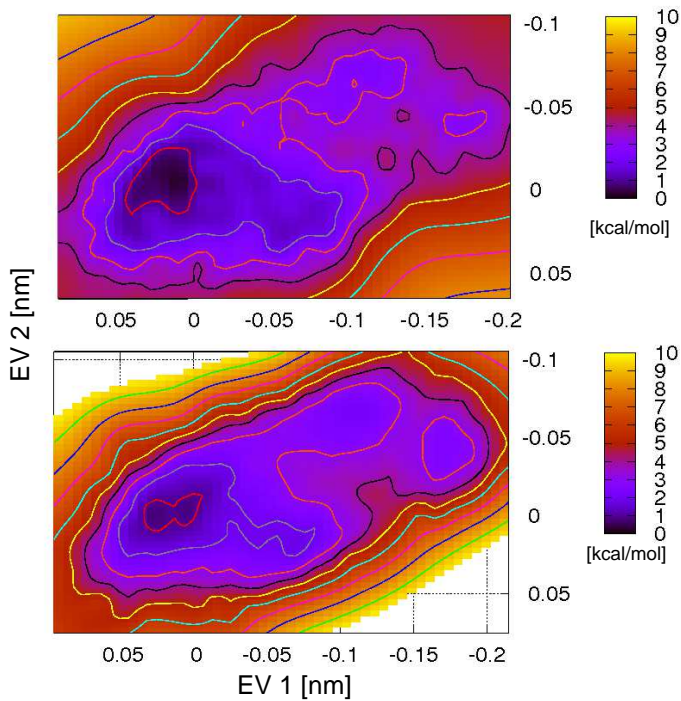}
\caption{
    \label{fig:comparison} Free energy landscape for the first two eigenvectors in the conventional metadynamics scheme (top) and in the histogram reweighted metadynamics scheme (bottom). The lines denote energy 
	differences
	of 0.5 kcal/mol.}
\end{figure}
Finally we compare the results of the histogram reweighted metadynamics simulation in the phase space of the eigenvectors to a conventional metadynamics simulation with 2000 hills.
As the landscapes are nearly identical, it can be concluded that our method is valid. Only slight deviations can be observed for the specific shape of the minina
due to the statistical error inherent in the original metadynamics variant \cite{Laio08}.
\subsection{Met-Enkephalin}
\begin{figure}
\includegraphics[width=0.5\textwidth]{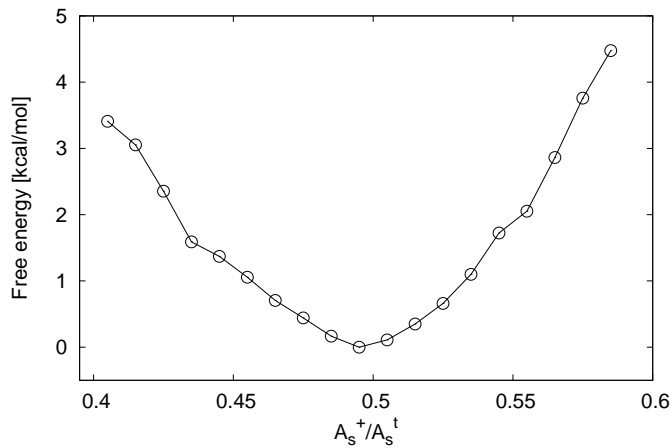}
\caption{ Free energy landscape for the ratio of the hydrophilic solvent accessible surface area $A_s^+$ to the total solvent accessible surface area $A_s^t$.
    \label{fig:eleven} }
\end{figure}

The Met-Enkephalin has attracted broad interest in its stable conformations \cite{Koca95,Perez92,Garcia02,Hansmann99} due to its biological importance.
Previous studies have shown that the main shape of the free energy landscape is given by a funnel \cite{Garcia02, Hansmann99,Evans2003} with a couple of stable conformations which
are separated by low energy barriers.
Regarding the biological function of the Met-Enkephalin, which is an opioid peptide that inhibits neurotransmitter release from the
appropriate opioid receptor, several receptors have to bind on the molecule which all require different stable conformations \cite{Hughes75,Kendrew94}.\\
The corresponding eigenvector analysis of the Met-Enkephalin illustrates that over 73\% of the
atomic
positional fluctuation have been captured by the first two eigenvectors.
Thus nearly all relevant structural transitions can be described by eigenvectors $1$ and $2$.\\
\begin{figure}
\includegraphics[width=0.5\textwidth]{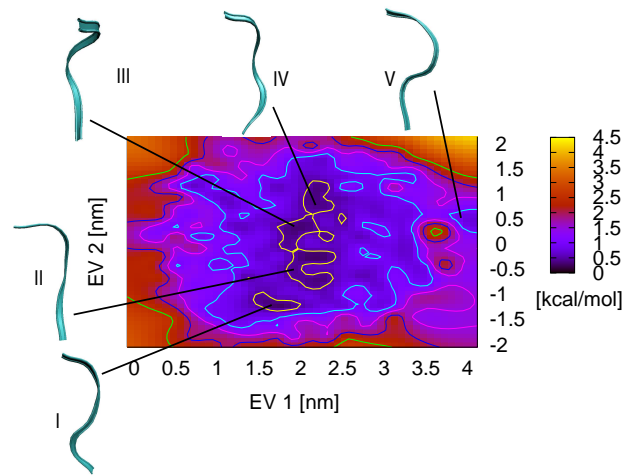}
\caption{ Free energy landscape for the eigenvectors $1$ and $2$ of the Met-Enkephalin. The lines correspond to energy differences of $0.5$ kcal/mol.
    \label{fig:nine} }	
\end{figure}
The corresponding free energy landscape at $T=300$ K is shown in Fig.~\ref{fig:nine}. A couple of stable minima and configurations can be found in the funnel-like landscape in agreement 
to other publications \cite{Garcia02,Hansmann99,Evans2003}.
The lines correspond to energy differences of $0.5$ kcal/mol. It is obvious that all energetic barriers are lower than $1.5$ kcal/mol.
Hence the transitions between the stable
configurations are not drastically hindered.
Regarding the conformations shown in Fig.~\ref{fig:nine}, differences in their form and shape can be observed reflecting the biological function of the
Met-Enkephalin \cite{Kendrew94}. The energetic flexibility in
the illustrated conformations finally explains the lack of a poor experimental convergence to a single
structure \cite{Graham92}.\\
Another interesting quantity is the solvent accessible surface area which allows the investigation of the importance of the solvent interactions for the Met-Enkephalin.
The influence of the hydrophilic solvent accessible surface area $A_s^+$ , which represents the influence of hydration effects on the stability of the peptide can be
investigated by the ratio to the total solvent accessible surface
area $A_s^t$ which is illustrated in Fig.~\ref{fig:eleven}. \\
\begin{figure}
\includegraphics[width=0.5\textwidth]{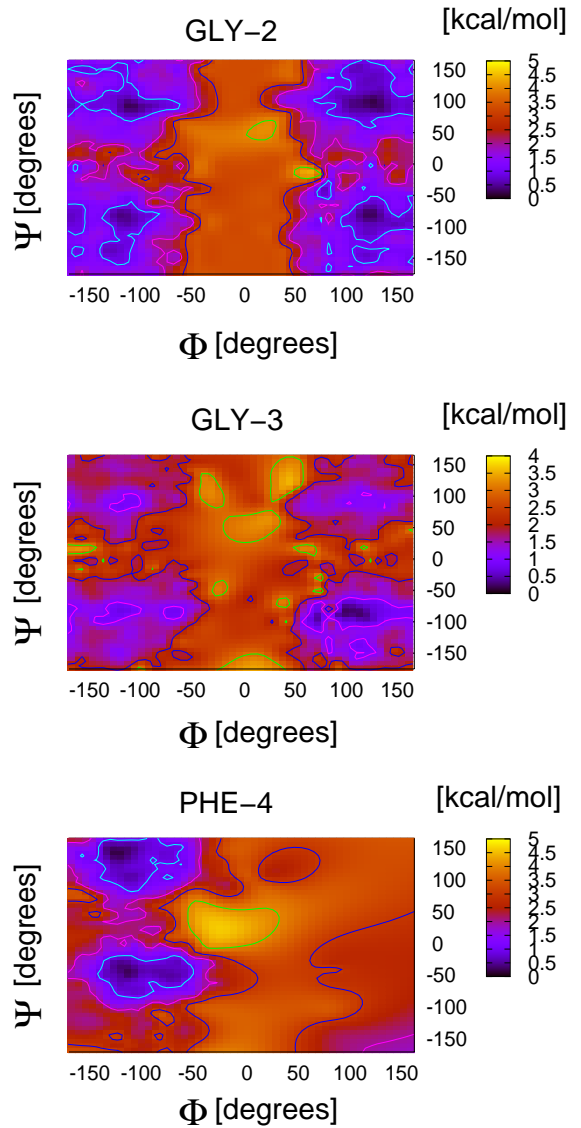}
\caption{ Free energy landscape in the Ramachandran plot for three residues. The lines correspond to energy differences of $1.0$ kcal/mol.
    \label{fig:twelve} }
\end{figure}
A large minimum can be found at a ratio of $A_s^+/A_s^t\approx 0.49$. It can be shown that nearly all stable configurations presented in Fig.~\ref{fig:nine} obey this
characteristic value for the ratio. Hence the formation of different stable conformations can be partly explained by hydration effects due to the chemical structure and
solubility of the Met-Enkephalin.
The rapid increase in the free energy of 4 kcal/mol for fluctuations $\delta(A_s^+/A_s^t)\approx 0.1$ avoids the appearance of drastic structure transitions around this
minimum.\\
The rigidity and flexibility of the Met-Enkephalin and its residues is finally illustrated in Fig.~\ref{fig:twelve} by a Ramachandran plot.
Here the residues GLY-2, GLY-3 and PHE-4 are investigated concerning their mechanical flexibility. All residues are able to form $\beta$-sheets while especially
the dihedral angles of PHE-4 prefer to visit the $\alpha$-helical conformations \cite{Garcia02,Cornell97,Head91}. Left-handed helical conformations can be found in
GLY-2 and GLY-3.
\section{Computational cost}
\begin{figure}
\includegraphics[width=0.5\textwidth]{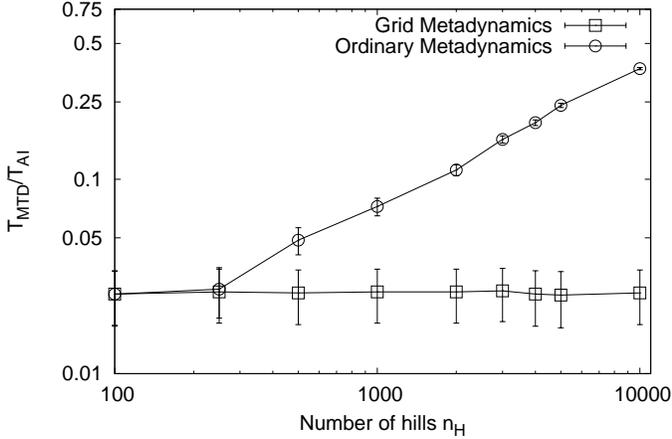}
\caption{
    \label{fig:alanintimes} Ratio for the time needed to calculate the metadynamics forces versus all interactions for one timestep in dependence of an increasing number of hills for the alanine dipeptide.}
\end{figure}

We finally have investigated the computational cost of the histogram reweighted metadynamics scheme in comparison to the ordinary metadynamics technique studied by the two dimensional 
test case of Eqn.~(\ref{eq:2d}).
\begin{figure}
\includegraphics[width=0.5\textwidth]{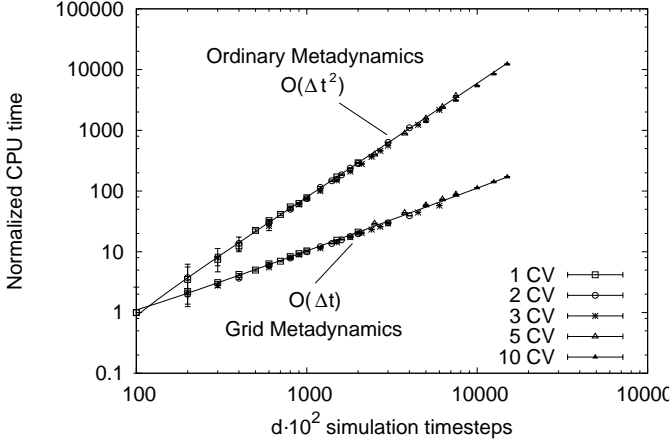}
\caption{\label{fig:times}
Normalized computational cost for the construction of the potential energy landscape of Eqn.~(\ref{eq:2d}) in the grid scheme in comparison to conventional metadynamics for different numbers of 
collective variables $d=1,2,3,5$ and $10$. 
}
\end{figure}
Thus we simulated the two dimensional test case by a conventional metadynamics scheme \cite{Laio08} where every hundredth step a gaussian hill of height 0.1$\epsilon$ with a width of 0.25$\sigma$ has been set.
The same values have also been used in the grid scheme. We compared the times needed for the construction of the potential energy landscape using the $x$ and $y$ direction as reaction coordinates and normalized them.\\ 
To analyze the influence of the number of collective variables $d$ we conducted simulations with $1,2,3,5$ and 10 collective variables.
It is obvious that the grid metadynamics
scheme scales as $\mathcal{O}(d t)$ whereas conventional metadynamics obeys a $\mathcal{O}(d t^2)$ behaviour (Fig.~\ref{fig:times}).
Although the program is not computationally 
optimized due to output-input data flow, 
it becomes clear that the overall time needed for the conventional metadynamics algorithm is increasing with the second power of simulation steps, respectively present hills. 
Hence it is evident that the number of hills is the dominating factor.
This leads us to the conclusion that our method can be used for a larger number of dimensions as well with a better scaling behaviour than the conventional scheme.\\
The ratio of the time used for the calculation of the metadynamics algorithm $T_{_{MTD}}$ compared to the time used for all interactions $T_{_{AI}}$ 
for an increasing number of apparent hills $n_{_{H}}$ is shown in Fig.~\ref{fig:alanintimes}.
All data points have been derived by the simulation of the alanine dipeptide.
It is evident that after 250 hills the calculational cost increases drastically in the ordinary metadynamics algorithm whereas for the grid technique the required time remains constant. 
Thus the grid variant of the metadynamics algorithm accelerates the 
computation of free energy landscapes in contrast to ordinary metadynamics considerably.\\
Nevertheless it can be assumed that 
the total time needed for the evaluation of the biasing forces in one timestep is negligible compared to the number of interactions especially for systems with many degrees of freedom.
Thus for a system with a large number of interactions, 
the cost for the evaluation of the metadynamics forces may decrease in comparison to the total time.\\ 
To investigate this situation and to derive an empirical formula, we studied the computational cost for the Met-Enkephalin and the alanine dipeptide. 
Hence we compared the total time which is required for the computation of the conventional metadynamics forces without a grid $\tau_{_{MTD}}$ to the total time of an unbiased run $\tau_{_{ub}}$.
\begin{figure}
\includegraphics[width=0.5\textwidth]{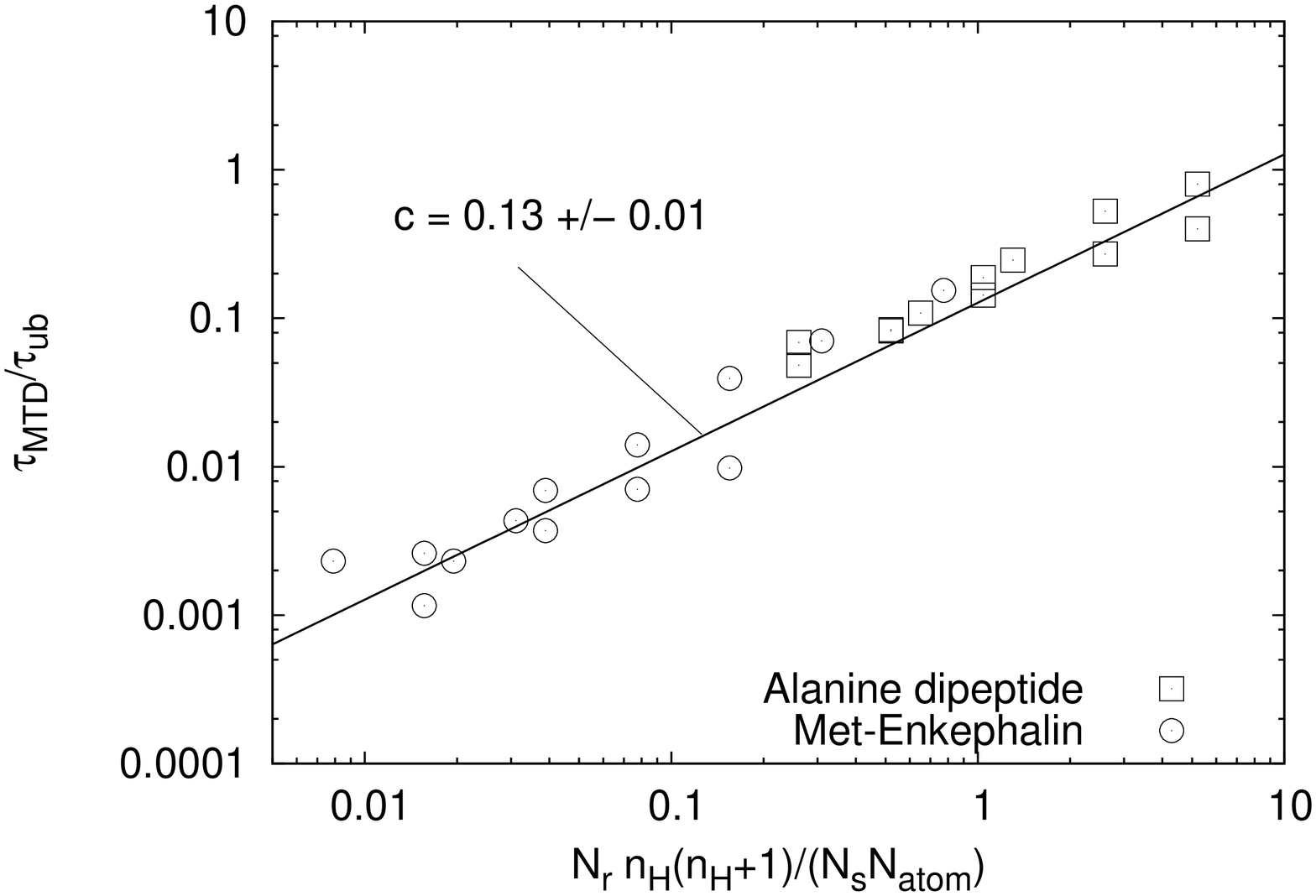}
\caption{Time needed for computation of the ordinary metadynamics forces $\tau_{_{MTD}}$ of Met-Enkephalin and alanine dipeptide 
	in comparison to the time for unbiased simulations $\tau_{_{ub}}$ for a varying number of hills $n_{_{H}}$ and number of relaxation steps $N_r$. 
	All values follow a linear relation with slope $c\sim 0.13$ and the number of collective variables was $d=2$.  
    \label{fig:thirteen} }
\end{figure} 
It can be assumed that the time for the calculation of the metadynamics forces depends on the number of present hills $n_{_{H}}$, the number of collective variables $d$ and the number of relaxation steps $N_r$ 
after that a new hill is deposited. In addition, the contributing number of atoms in the collective variable may also influence the computation time. For reasons of simplicity we assume that this time
is mainly determined by the evaluation of $\mathcal{G}(x-x_{_{G,i}})$ (Eqn.~(\ref{eq:weight})) where all other effects are negligible.
Thus for much larger systems this factor may become size-dependent.\\
In summary the total time needed for the evaluation of the ordinary metadynamics forces in a simulation can be written as 
$\tau_{_{MTD}}\sim d N_r\sum_{j=1}^{n_{_{H}}}j \sim d N_r n_{_{H}} n_{_{H}} (n_{_{H}}+1)\delta t_{_{MTD}}/2$ where $\delta t_{_{MTD}}$ denotes the time needed for the evaluation of a single hill. Additionally it is further
assumed that the simulation has finished after $n_{_{H}}$ hills have been set which is given by the relation $n_{_{H}}=N_{_{s}}/N_{_{r}}$.\\ 
Furthermore the total time for an unbiased simulation $\tau_{_{ub}}$ is nearly proportional to the number of atoms in the system $N_{_{atom}}$ \cite{Smit} and the number of simulation timesteps $N_{_{s}}$ 
which yields $\tau_{_{ub}}\sim N_{_{atom}}N_{_{s}} \delta t_{_{ub}}$ with the average time $\delta t_{_{ub}}$ needed for the calculation of a single unbiased force. This gives the ratio
\begin{equation}
\frac{\tau_{_{MTD}}}{\tau_{_{ub}}} \sim \frac{ d N_r n_{_{H}} (n_{_{H}}+1)}{2 N_{_{s}} N_{_{atom}}} \frac{\delta t_{_{MTD}}}{\delta t_{_{ub}}} 
\sim \frac{ d n_{_{H}}}{2N_{_{atom}}} \frac{\delta t_{_{MTD}}}{\delta t_{_{ub}}}
\label{eq:slope}
\end{equation}
where $\delta t_{_{MTD}}/\delta t_{_{ub}}$ can be empirically determined. It has to be mentioned that the ratio is strongly dependent on the algorithms which are used for the evaluation of the forces 
but not on the processor speed.\\ 
We have used varying relaxation timesteps of 
$N_r=1 - 1000$, different numbers of hills $n_{_{H}}=50 - 10000$ and the number of atoms was $N_{_{atom}}=3231$ for the Met-Enkephalin with water and  $N_{_{atom}}=961$ for the alanine dipeptide plus water
whereas $N_s$ varies between 100 and 100000.\\
The values for different ordinary metadynamics runs in comparison to unbiased runs for the Met-Enkephalin and the alanine dipeptide are shown in Fig.~\ref{fig:thirteen}. It is evident that all values 
follow Eqn.~(\ref{eq:slope}) with a proportionality factor of $c=0.13 \pm 0.01$. Hence for a large number of present hills, 
the ordinary metadynamics method drastically increases computation time. Thus as a general remark, the original method becomes computationally very expensive if the 
number of hills is larger than the number of atoms. This is often given for systems with implicit solvent models or complex and large energy landscapes.\\ 
As we have shown before, the grid based technique obeys a linear $\mathcal{O}(d t)$ behaviour, such that the ratio between the time needed for an ordinary metadynamics algorithm in comparison to the grid technique
also grows in accordance to Eqn.~(\ref{eq:slope}). 
Especially for ordinary metadynamics simulations where the free energy landscape is not refined by histogram 
reweighting procedures the amount of settled hills often increases the number of atoms which strongly suggests the use of grid based techniques. This is mostly important for classical Molecular Dynamics simulations
for which our method is more intended in contrast to ab-intio methods. The evaluation of the electronic motion is here the main factor \cite{Leach01} 
dominating most of the computation time such that the evaluation of the metadynamics potential can be neglected.
\section{Summary and conclusion}
We have presented an efficient and simple method for the calculation of free energy landscapes. The technique is applicable for a broad range of different systems. The basic principles are the construction of
a potential landscape on a predefined grid in an initial simulation run which is used as a biasing potential in the
final simulations. The corresponding probability distributions of the biased
simulations are reweighted by the WHAM procedure whose usage avoids specific
drawbacks of the ordinary metadynamics algorithm like large error tolerances \cite{Babin06}.\\
Another advantage is the easy implementation due to its simple methodology in software packages like GROMACS.
In addition we have presented the application of a projection scheme which allows to transform the energy landscape of certain collective variables to further reaction coordinates
without any additional simulation effort. The specific choice of biasing the
first eigenvectors of the system achieves a good activation of the whole system.\\
The history-dependent potential is short-ranged in contrast to the
ordinary Metadynamics scheme \cite{Laio02,Laio08} and serves as a pure
biasing potential. The fine resolution of the free energy landscape is
achieved by histogram reweighting techniques of the corresponding probability
distributions. A further advantage of the method is the opportunity to tune the resolution of the
landscape even after the simulations were finished.\\ 
Furthermore we have explicitly 
shown that a grid-based technique scales with $\mathcal{O}(dt)$ in contrast to the conventional metadynamics scheme ($\mathcal{O}(dt^2)$), 
where $t$ denotes the simulation time and $d$ the number of applied collective variables. Hence, compared to the ordinary metadynamics algorithm a grid technique is more preferable due to a decreased simulation time.
This has been shown by the computational cost for the alanine dipeptide and the Met-Enkephalin. Via a linear relation we were able to show that the computational cost of the ordinary metadynamics method 
in comparison to the grid technique can be predicted by an empirical formula. Thus the grid method is preferable for systems where the number of settled hills exceeds the number of all atoms in the system. 
As we have discussed, this fact is given for implicit solvation models as well as very large energy landscapes.\\  
In addition our method has been tested for the peptides alanine dipeptide and
Met-Enkephalin. The results are in good agreement to the literature.
We have further shown that the energy landscape of the Met-Enkephalin is funnel-like with many different conformations at several energetic minima.
The similarity between these conformations can be
illustrated by a characteristic ratio of the hydrophilic solvent accessible surface area in comparison to the total solvent accessible area which results in a characteristic
minimum around $A_s^+/A_s^t \approx 0.49$.
The flexibility of the Met-Enkephalin is illustrated by plotting the dihedral
angles of each residue in a Ramachandran plot.\\
Additionally we have shown that the method allows to identify the stable
conformations of the alanine dipeptide with good accuracy of the free energy
landscape. Following \cite{Laio08,Babin06}, where it was recognized that the functionality
of a biasing potential does not depend on its specific shape, we have
further validated our technique by the explicit calculation of two well-defined test
cases such that it can be used as an additional variation of the existing metadynamics methods.
\section{Acknowledgements}
We thank Clotilde Cucinotta, Wolfgang Wenzel, Hans Behringer, J. Arne M{\"u}ller and Marc Hageb{\"o}lling for fruitful and enlightening discussions. 
The authors also have to thank Vojtech Spiwok for using parts of his metadynamics source code available on his web page \cite{Vojtech}.
Financial support by the Deutsche Forschungsgemeinschaft (DFG) through the transregional collaborative research center
TRR 61 is gratefully acknowledged.


\end{document}